\documentclass[twocolumn,twocolappendix,tighten]{aastex631}

\received{2022 June 17}
\revised{2022 August 1}
\accepted{2022 August 1}
\published{date}

\graphicspath{{./}{figures/}}

\newcommand{\rsun}{R_\odot}
\newcommand{\msun}{M_\odot}

\newcommand{\cc}{{\rm cm^{-3}}}

\shorttitle{Magnetic field growth around the Pop III star}
\shortauthors{Hirano and Machida}

\begin{document}

\title{Exponentially amplified magnetic field eliminates disk fragmentation \\around the Population III protostar}

\correspondingauthor{Shingo Hirano}
\email{hirano@astron.s.u-tokyo.ac.jp}

\author[0000-0002-4317-767X]{Shingo Hirano}
\affiliation{Department of Astronomy, School of Science, University of Tokyo, Tokyo 113-0033, Japan}
\affiliation{Department of Earth and Planetary Sciences, Faculty of Science, Kyushu University, Fukuoka 819-0395, Japan}

\author[0000-0002-0963-0872]{Masahiro N. Machida}
\affiliation{Department of Earth and Planetary Sciences, Faculty of Science, Kyushu University, Fukuoka 819-0395, Japan}

\begin{abstract}
One critical remaining issue to unclear the initial mass function of the first (Population III) stars is the final fate of secondary protostars formed in the accretion disk, specifically whether they merge or survive.
We focus on the magnetic effects on the first star formation under the cosmological magnetic field.
We perform a suite of ideal magnetohydrodynamic simulations until 1000 years after the first protostar formation.
Instead of the sink particle technique, we employ a stiff equation of state approach to represent the magnetic field structure connecting to protostars.
Ten years after the first protostar formation in the cloud initialized with $B_0 = 10^{-20}$\,G at $n_0 = 10^4\,\cc$, the magnetic field strength around protostars amplifies from pico- to kilo-gauss, which is the same strength as the present-day star.
The magnetic field rapidly winds up since the gas in the vicinity of the protostar ($\leq\!10$\,au) has undergone several tens orbital rotations in the first decade after protostar formation.
As the mass accretion progresses, the vital magnetic field region extends outward, and the magnetic braking eliminates fragmentation of the disk that would form in the unmagnetized model.
On the other hand, assuming a gas cloud with small angular momentum, this amplification might not work because the rotation would be slower.
However, disk fragmentation would not occur in that case.
We conclude that the exponential amplification of the cosmological magnetic field strength, about $10^{-18}$\,G, eliminates disk fragmentation around the Population III protostars.
\end{abstract}


\section{Introduction} \label{sec:intro}

One of the significant challenges in modern cosmology is the formation process of the first generation of stars, the so-called Population III (Pop III) stars.
They influence all subsequent star and galaxy evolution in the early Universe through their input of ionizing radiation and heavy chemical elements, depending on the final fates of Pop III stars \citep{yoon12}.
There have been no direct observations yet, but the nature of the first stars has been elucidated by theoretical studies, in particular with numerical simulations of increasing physical realism \citep[for recent reviews,][]{greif15}.
Furthermore, several indirect constraints exhibit the imprint of the first stars: $<\!0.8\,\msun$ low-mass stars capable of surviving to the date \citep[e.g.,][]{magg18}, about $100\,\msun$ massive binaries which can be a promising progenitor of BH-BH (blackhole) merger like the gravitational wave sources \cite[e.g.,][]{kinugawa14}, and $\sim\!10^5\,\msun$ supermassive stars which can leave massive seed BHs of the high-$z$ supermassive black holes \citep[SMBHs; e.g.,][]{inayoshi20}.
There is a need to update the theoretical model for the formation and evolution of the first stars to predict their observational signature in light of the upcoming suite of next-generation telescopes.

One of the key unresolved issues in the first star formation theory is the efficiency of magnetic effects (e.g., magnetic braking).
Previous studies have identified several effects assuming primordial star-forming gas clouds have strong magnetic fields: delaying the gas contraction to the host DM minihalo and the first star formation \citep[e.g.,][]{koh21}, preventing disk fragmentation with efficient angular momentum transport due to the magnetic field \citep{machida13,sadanari21}, and reducing the protostellar rotation degree which can also control the final fate of Pop III stars \citep{hirano18}.
However, it is known that the primordial magnetic field of the universe \citep[$10^{-18}$\,G;][]{ichiki06} is extremely weak compared to the magnetic field of nearby star-forming regions ($\sim\!10^{-6}$\,G).
The magnetic field amplification by the flux freezing during the cloud collapse phase, $B \propto n^{2/3}$, is insufficient to provide the magnetic field strength to affect the first star formation.

The small-scale turbulent dynamo can lead further amplification \citep[summarized in][]{mckee20}.
Cosmological magnetohydrodynamical (MHD) simulations provide power-law fits to their results, to be compared to the flux-freezing expression, specifically $B \propto n^{0.83}$ \citep{federrath11} and $B \propto n^{0.89}$ \citep{turk12}.
However, the amplification level via the turbulent small-scale dynamo depends on the numerical resolution \citep{sur10,sur12}.
Recent MHD simulation \citep{stacy22} showed that the small-scale dynamo could contribute only one or two orders of magnitude to the magnetic field amplification during gas cloud contraction.
Hence, most of the amplification comes from compressional flux-freezing.

Recently, \cite{hirano21} reported a new mechanism of magnetic field amplification during the protostellar accretion phase in the primordial atomic-hydrogen (H) cooling gas clouds.
Many fragments of a gravitationally unstable gas cloud amplify the magnetic field due to the rotational motion and form a vital magnetic field region around the protostar.
The simulations in \cite{hirano21} adopted the stiff equation of state (EOS) technique then we could calculate the coupling between the high-density region and the magnetic field, which is essential in reproducing this amplification mechanism.
However, MHD simulations of the first star formation replaced the dense region by the sink particle \citep[$n_{\rm sink} \sim 10^{13}\,\cc$;][]{sharda20,sharda21,stacy22} which results in the loss of the magnetic field connection from the dense region.
Therefore, it is not uncovered whether a similar amplification occurs in the primordial molecular hydrogen (H$_2$) cooling gas clouds in such simulations.
Our previous simulations \citep{machida13} assumed a relatively strong magnetic field ($10^{-10}$--$10^{-5}$\,G at $n = 10^4\,\cc$) and did not consider a cosmological weak magnetic field as an initial condition.

We performed three-dimensional ideal MHD simulations of the primordial star formation using the stiff-EOS technique.
We find that exponential magnetic field amplification occurs in the vicinity of the protostar in the first three years after the first Pop III protostar formation and that the field-amplified region completely suppresses disk fragmentation.
This letter introduces this new exponential magnetic field amplification mechanism and substantial expansion of the amplified region.
In the following Paper II, we will discuss the effects of model parameters on the first star formation in detail.

\begin{figure}[t]
\begin{center}
\includegraphics[width=1.0\columnwidth]{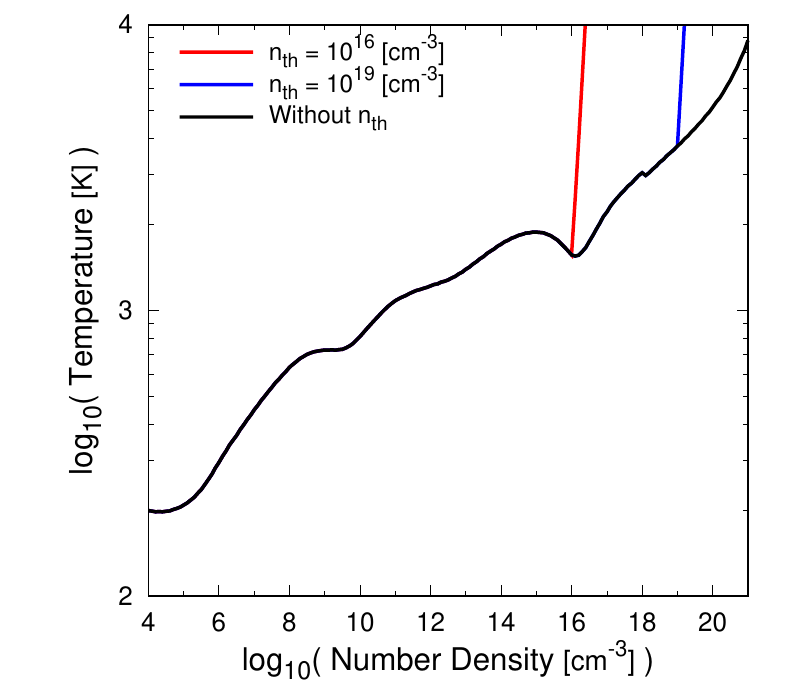}
\end{center}
\caption{
Thermal evolution models of the zero-metallicity star-forming clouds as a function of the gas number density.
The black line is the base model theoretically obtained by the chemical reaction calculation \citep{omukai08,machida15}.
The colored lines, the adopted models in this study, are variants of the black line using the stiff equation of state technique with threshold density $n_{\rm th} = 10^{16}\,\cc$ (red) and $10^{19}\,\cc$ (blue), respectively.
}
\label{f1}
\end{figure}

\section{Methods} \label{sec:methods}

\subsection{Numerical methodology}

We solve the ideal MHD equations with the barotropic equation of state (EOS).
Note that non-ideal MHD effects are not effective in primordial gas clouds \citep[e.g.,][]{higuchi18}.
To represent the thermal evolution in the zero-metallicity cloud, we adopt the EOS table based on a chemical reaction calculation \citep{omukai08}.
Instead of the sink particle technique, we employ the stiff-EOS approach to represent the magnetic field structure connecting to the dense gas region which is established in our past works \citep{machida13,machida14,hirano17,hirano21,susa19}.
Figure~\ref{f1} shows the resultant EOS tables.
We adopt two threshold densities $n_{\rm th}$ to study the early amplification and later expansion of the magnetic field, respectively: (1) simulations until $t_{\rm ps} = 100$\,yr with $n_{\rm th} = 10^{19}\,\cc$ which reproduce hydrostatic cores whose radius is consistent with the mass-radius relation of an accreting primordial protostar \citep{hosokawa10} and (2) until $t_{\rm ps} = 1000$\,yr with $n_{\rm th} = 10^{16}\,\cc$.
We define the epoch of the first protostar formation ($t_{\rm ps} = 0$\,yr) when the gas number density firstly reached the threshold density ($n_{\rm max} = n_{\rm th}$).

We use the nested grid code \citep{machida15}, in which the rectangular grids of ($n_{\rm x}$, $n_{\rm y}$, $n_{\rm z}$) = ($256$, $256$, $32$) are superimposed.
The base grid has the box size $L(0) = 9.83 \times 10^5$\,au and the cell size $h(0) = 3.84 \times 10^3$\,au.
A new finer grid is generated to resolve the Jeans wavelength at least 32 cells.
The maximum grid levels and the finest cell sizes are $l = 17$ and $h(17) = 0.0293$\,au for runs with $n_{\rm th} = 10^{19}\,\cc$ whereas $l = 14$ and $h(14) = 0.234$\,au for $n_{\rm th} = 10^{16}\,\cc$, respectively.

\subsection{Initial Condition}
\label{sec:initial}
The initial cloud has a enhanced Bonner-Ebert (BE) density profile $n(r) = f \cdot n_{\rm BE}(r)$ with a enhanced factor $f = 1.6$ to promote the cloud contraction.
The initial central density is $f \cdot n_{\rm BE}(r = 0) = f \cdot 10^4\,\cc$.
The mass and size of the initial cloud are $M_{\rm cl,0} = 4.83 \times 10^{3}\,\msun$ and $R_{\rm cl,0} = 2.38$\,pc, respectively.
A rigid rotation of $\Omega_0 = 1.31 \times 10^{-14}\,{\rm s^{-1}}$ is imposed.
With these settings, the thermal and rotational energies to the gravitational energy of the initial cloud are $\alpha_0 = 0.533$ and $\beta_0 = 0.0209$, respectively.
We do not include turbulence and do not consider a small-scale dynamo \citep[e.g.,][]{sur10,mckee20}, because we only consider very weak fields which are significantly amplified by the rotation motion of protostars (see \S\ref{sec:dis} for details).

\subsection{Model Parameter}

We impose a uniform magnetic field $B_0$ with the same direction as the initial cloud's rotation axis in the whole computational domain.
We examine the parameter dependence of the first star formation on $B_0 = 0$, $10^{-20}$, $10^{-15}$, and $10^{-10}$\,G (labels as B00, B20, B15, and B10).
We adopt B20 as the fiducial model in this study because $B_0 = 10^{-20}$\,G is lower than the cosmological value $\sim\!10^{-18}$\,G \citep{ichiki06}.
If we can show that magnetic fields affect first star formation in this model, we can prove that the first stars cannot avoid the effects of magnetic fields.
By comparing B20 with B00, we examine the magnetic effect on the first star formation.
By comparing B20 with B15 and B10, we study the necessity of the early amplification of the magnetic field strength before the star-forming cloud formation.

\section{Results} \label{sec:res}

This section shows the results of MHD simulations with different initial magnetic field strengths (B00, B20, B15, and B10).
We run four models under different threshold densities until $t_{\rm ps} = 100$\,yr ($n_{\rm th} = 10^{19}\,\cc$) and $1000$\,yr ($n_{\rm th} = 10^{16}\,\cc$).
Since the calculation results among models with different resolutions converged outside the resolution limit\footnote{We find no resolution-dependence effects like as MRI in our simulations.}, this section shows the combined results of the first $100$ years under $n_{\rm th} = 10^{19}\,\cc$ and the latter $900$ years under $n_{\rm th} = 10^{16}\,\cc$.

\subsection{Fiducial model}

\begin{figure*}[t]
\begin{center}
\includegraphics[width=1.0\linewidth]{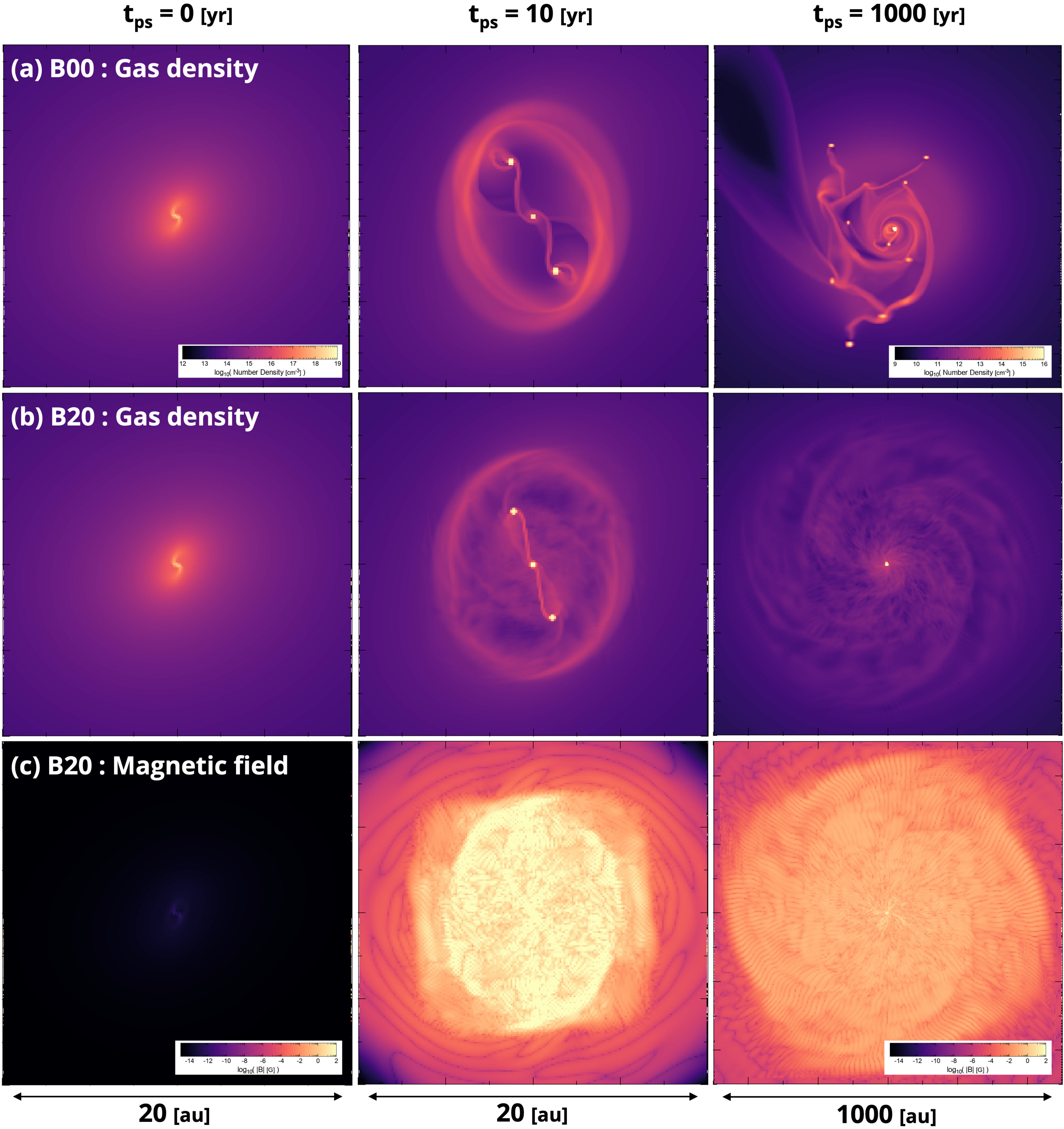}
\end{center}
\caption{
Cross-sectional view on the $z = 0$ plane around the most massive protostar at $t_{\rm ps} = 0$, $10$, and $1000$\,yr after the first protostar formation from left to right.
Panels: (top) gas number density in model B00, (middle) gas number density in model B20, and (bottom) absolute magnetic field strength in model B20.
The box sizes are $20$\,au in left and middle panels and $1000$\,au in right panels, respectively.
}
\label{f2}
\end{figure*}

Figure~\ref{f2} compares the simulation results of the fiducial model (B20) and the unmagnetized model (B00) during the first 1000 years of the protostar accretion phase.
At the birth time of the first protostar (left panels), the density structure around the protostar is identical because the magnetic field strength in the vicinity of the protostar is too weak (pico-gauss $= 10^{-12}$\,G at most) to affect the collapsing gas cloud.
However, after a decade (middle column in Figure~\ref{f2}), the magnetic field strength on the primary protostar surface ($\sim\!30\,\rsun$) amplifies to kilo-gauss \citep[similar to the Pop I protostars;][]{johns-krull07}, and this strong ``seed'' field amplifies the surrounding field in the region of $10$\,au radius.
Within this strong magnetic field region, the density and velocity structure of accretion gas are affected by the magnetic field.
After 1000 years (right column in Figure~\ref{f2}), the amplified magnetic field region extends to a radius of about $500$\,au and multiple protostars that appeared in the unmagnetized model have disappeared in the fiducial model.
The global spiral structure of gas appears inside the amplified region due to the angular momentum transport by magnetic braking that allows accretion to proceed efficiently.

\begin{figure*}[t]
\begin{center}
\includegraphics[width=1.0\columnwidth]{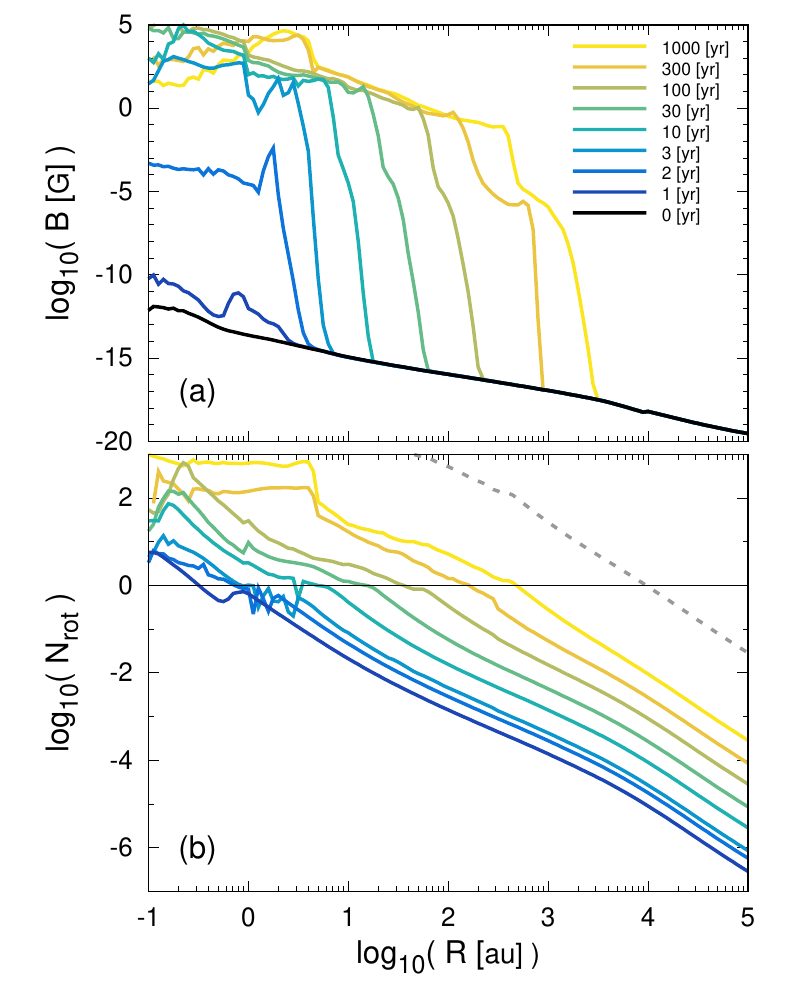} \includegraphics[width=1.0\columnwidth]{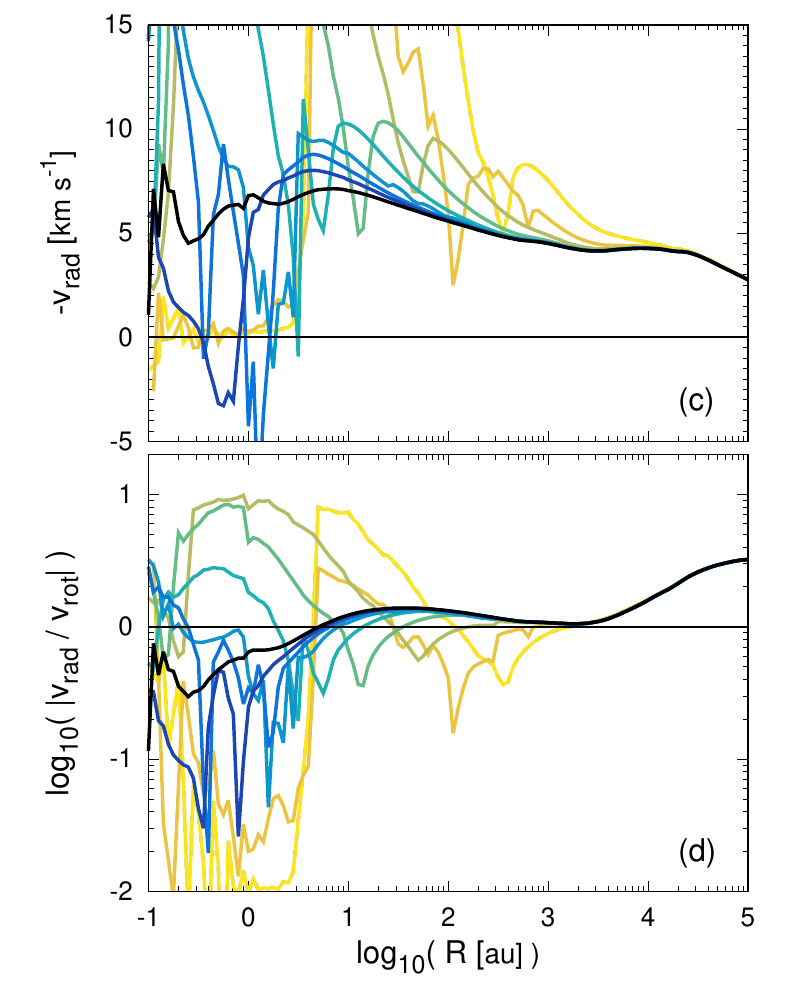}
\end{center}
\caption{
Radial profiles for model B20 at $t_{\rm ps} = 0$, $1$, $2$, $3$, $10$, $30$, $100$, $300$, and $1000$\,yr after the first protostar formation.
Panels: (a) magnetic field strength, (b) number of orbital rotations at $t_{\rm ps}$, $N_{\rm rot} = (v_{\rm rot} t_{\rm ps})/(2 \pi R)$, (c) radial velocity, and (d) absolute value of the ratio of the radial velocity to the rotational velocity.
The line for $t_{\rm ps} = 0$\,yr is not plotted in panel (b) since $N_{\rm rot} = 0$ for every $R$ at $t_{\rm ps} = 0$\,yr.
The dotted line in panel (b) represents the expected $N_{\rm rot}$ at $t_{\rm ps} = 10^5$\,yr when the first star ends its accretion phase \citep[fig~1 in][]{hirano17}, by using the radial profile of rotational velocity at $t_{\rm ps} = 1000$\,yr.
The horizontal lines indicate $N_{\rm rot} = 1$ in panel (b), $-v_{\rm rad} = 0$ in panel (c), and $\vert v_{\rm rad} \vert = \vert v_{\rm rot} \vert$ in panel (d).
}
\label{f3}
\end{figure*}

The origin of this exponential magnetic field amplification from $10^{-12}$\,G to $10^{3}$\,G nearby protostars is rapid rotational motion capable of winding up magnetic fields.
The magnetic field strength in the vicinity of the protostar has sufficiently amplified in the first three years after protostar formation (Figure~\ref{f3}a).
In the region of $\le\,10$\,au, the number of orbital rotations exceeds one at $t_{\rm ps} = 0$\,yr and reaches several dozen at $t_{\rm ps} = 3$\,yr (Figure~\ref{f3}b).
The magnetic field amplification region then widens over time, and its arrival radius equals the radius at which the orbital rotation rate exceeds one, $N_{\rm rot} = 1$.
Figure~\ref{f3}(c) shows the negative radial velocity $-v_{\rm rad}$ and indicates that the amplified field cannot significantly impede the gas accretion to the central region.
Figure~\ref{f3}(d) plots the ratio of the radial velocity to the rotational (or azimuthal) velocity.
The figure shows that the gas falls toward the center while rotating.
These figures mean that gravitational energy is efficiently converted into magnetic energy through kinetic (or rotational) energy after the first protostar formation.
We expect that the amplified magnetic field region could spread about $10^4$\,au $\sim\!0.05$\,pc, inside which the total gas mass is about $\sim 500\,\msun$ in this model, until the end of the accretion phase of the first stars (about $10^5$\,yr; dotted line).

\begin{figure}[t]
\begin{center}
\includegraphics[width=1.0\linewidth]{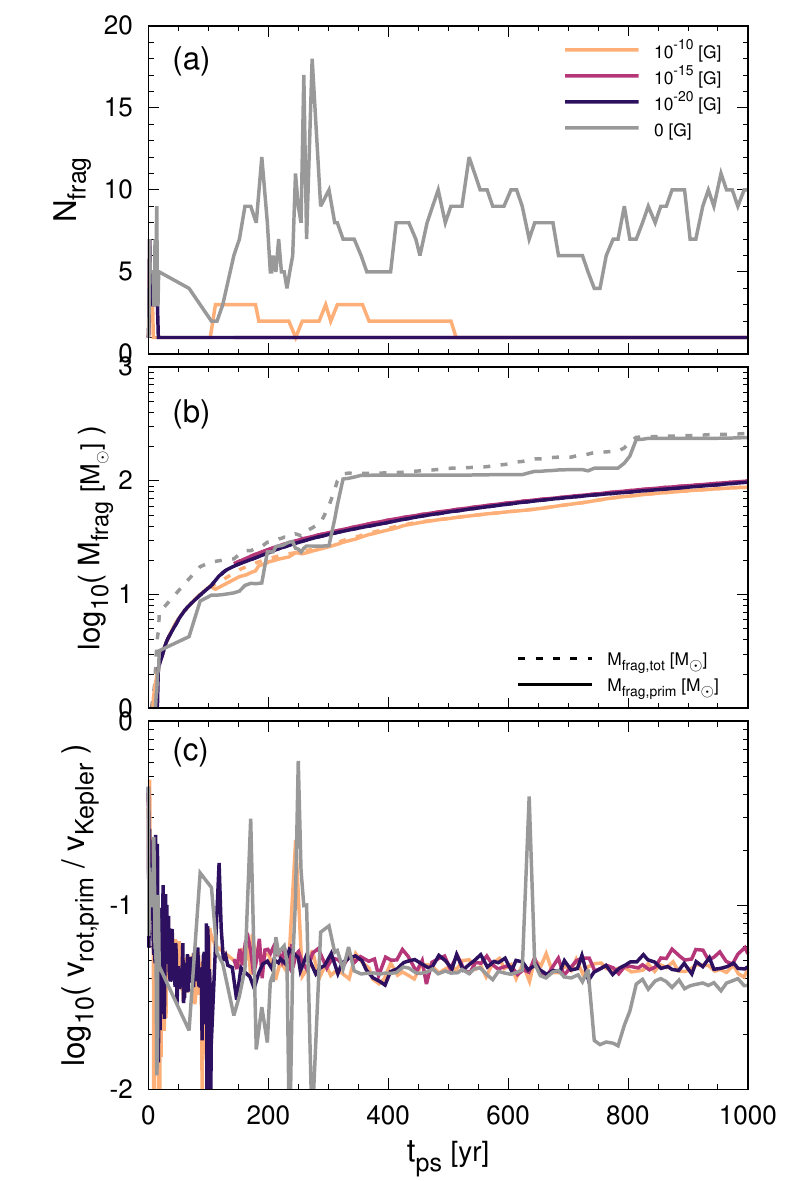}
\end{center}
\caption{
Time evolution of the protostar properties for models B00 (gray), B20 (blue), B15 (red), and B10 (yellow).
Panels: (a) number of protostars, (b) total mass of protostars $M_{\rm frag,tot}$ and mass of the most massive (primary) protostar $M_{\rm frag,prim}$, and (c) ratio of rotational velocity to the Keplerian velocity of the primary protostar.
}
\label{f4}
\end{figure}

\begin{figure*}[t]
\begin{center}
\includegraphics[width=1.0\linewidth]{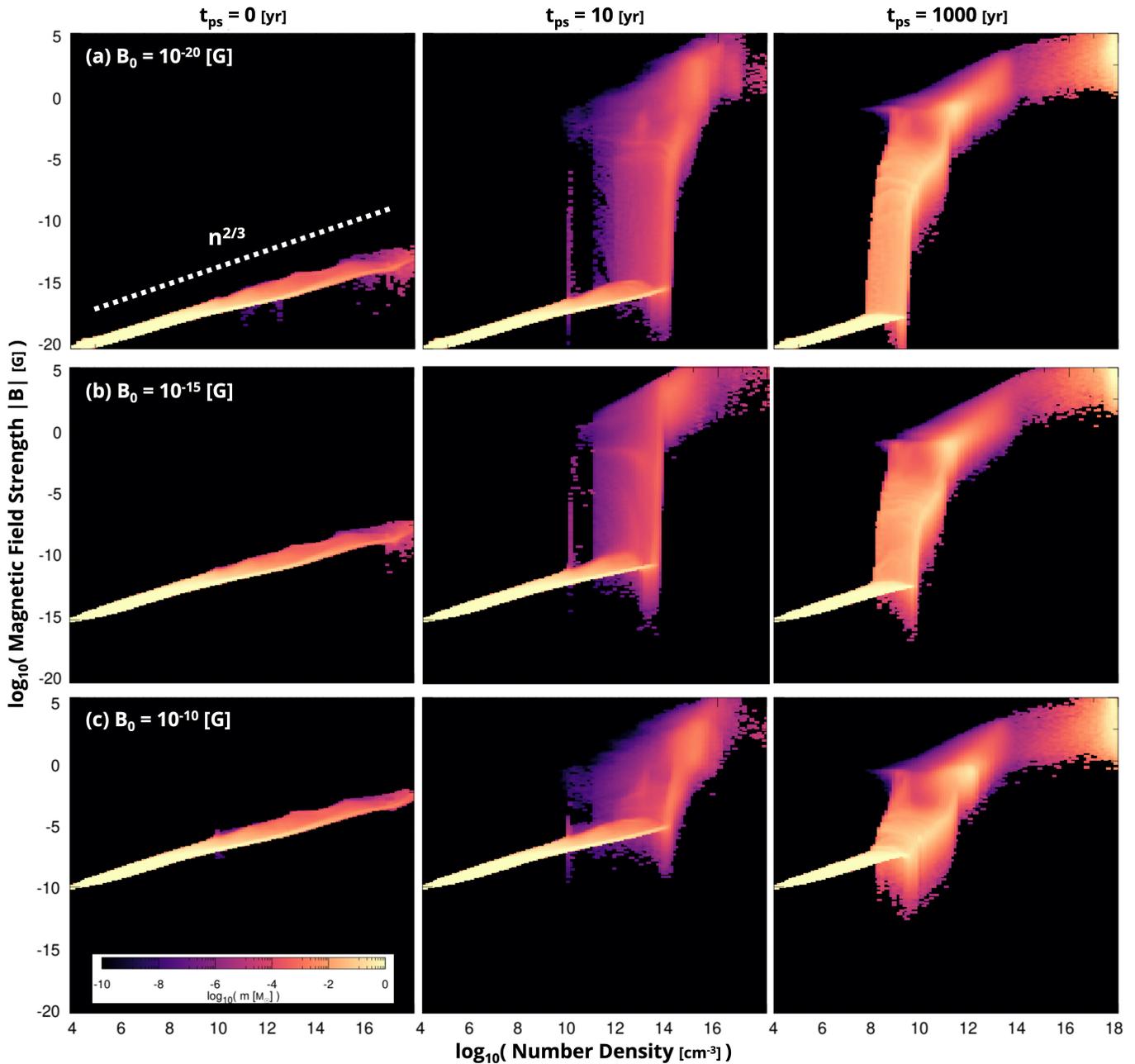}
\end{center}
\caption{
Phase diagrams of the absolute magnetic field strength for models B20, B15, and B10 (top to bottom) at $t_{\rm ps} = 0$, $10$, and $1000$\,yr after the first protostar formation (left to right).
The white dotted line in the left top panel shows the power law $B \propto n^{2/3}$ through flux freezing during the cloud compression.
}
\label{f5}
\end{figure*}

How does this exponentially amplifying magnetic field affect the formation process of the first stars?
The amplified magnetic field eliminates fragmentation of the gravitationally unstable accretion disk, and a single protostar forms at the center of the cloud (Figure~\ref{f4}a).
On the other hand, the stellar masses are not different between magnetized and unmagnetized models (Figure~\ref{f4}b).
In any case, the fragments born in the vicinity of the protostar will merge immediately.
The rotation velocities, the second important parameter of the stellar evolution theory, are nearly constant, $\sim\!0.05$ times the Keplerian velocity, regardless of the models and evolution time (Figure~\ref{f4}c).
Since the rotational degree is low, it seems reasonable to adopt a non-rotational model for the stellar evolution \cite[e.g.,][]{yoon12}.

\subsection{Dependence on the initial B-field strength}

We simulated two comparison models (B15 and B10) with higher initial magnetic field strengths than the fiducial model to examine the effects of other amplification mechanisms, which do not appear in our simulations.
The evolution of collapsing gas cloud is almost similar among the three models until $t_{\rm ps} = 0$\,yr except for the magnetic field strength distribution.
Figure~\ref{f5} presents that, in both cases, exponential magnetic field amplification occurs immediately after $t_{\rm ps} = 0$\,yr, similar to the fiducial model.
Because the expansion of the amplified magnetic field region completely prevents disk fragmentation, all magnetized models show the same results: the formation of a single first star (Figure~\ref{f4}a).

\section{Discussion} \label{sec:dis}

The magnetic field amplification after protostar formation proceeds in the following steps:
(1) The rotational motion of protostars amplifies the small magnetic field around them.
(2) The amplification rate of the magnetic field in the surrounding region increases according to the induction equation, $\partial B / \partial t = \nabla \times ( {\bf v \times B})$.
(3) The rotation-dominated region gradually extends outward, where the magnetic field is amplified by mechanism (1).
The MHD simulations adopting the sink particle technique cannot reproduce the ``seed'' magnetic field amplification around the protostar and the following propagation outwards because the rotation of the high-density region does not couple with the magnetic field.

The high accretion rate in the atomic hydrogen (H) cooling halo causes many fragments, which amplify the magnetic field due to the rotation \citep{hirano21}.
In the molecular hydrogen (H$_2$) cooling halo investigated in this study, the disk fragments appear only at the initial stage but soon merge into the primary protostar.
The orbital rotation around the protostar alone can amplify the magnetic field without further fragmentation.
This amplification mechanism is unique to star formation in the early Universe because it does not occur in nearby star-forming regions where the magnetic field saturates before the protostar accretion phase.
Thus, the feedback cannot be ignored under a strong magnetic field environment.

This magnetic field amplification prohibits fragmentation of accretion disk.
If the star-forming gas cloud has a sufficient rotational degree, a gravitationally unstable accretion disk forms and fragments, but the amplified magnetic field immediately suppresses disk fragmentation.
Conversely, if the rotation of the gas cloud is weak, the amplification of the magnetic field by the rotation is less efficient.
In this case, a gravitationally unstable accretion disk can not form and disk fragmentation does not occur.
In either case, a single first star remains.

We describe some caution about the rotational amplification of the magnetic field.
Recent studies have suggested turbulence as an amplification mechanism of magnetic field and shown that disk fragmentation is not significantly suppressed in turbulent environments \citep[e.g.,][]{sharda21,Prole22}.
We did not consider turbulence in this study (\S\ref{sec:initial}).
The rotational amplification mechanism may not be effective in highly turbulent environments because the turbulent reconnection (or reconnection diffusion) breaks the coupling between the magnetic field and gas (or fluid motion) even in ideal MHD calculations \cite[e.g.,][]{lazarian99,lazarian20}.
Thus, the existence of (strong) turbulence may significantly change our results, which will be investigated in our future paper.
In addition, the rotational amplification would not be efficient when magnetic dissipation, such as ambipolar diffusion and ohmic dissipation, is effective, and the amplified field significantly dissipates.
As shown in \cite{higuchi18}, the ambipolar diffusion becomes effective in the high-density region ($n\gtrsim 10^{12}\cc$) when the magnetic field strength exceeds $B \gtrsim 0.1 - 1$\,kG.
Thus, we need to consider the ambipolar diffusion in a further evolutionary stage.

Next, we discuss the amplification of the magnetic field and the treatment of protostars.
We have used the stiff-EOS technique instead of the sink particle technique, as in our  previous studies \citep[e.g.,][]{machida13,machida15,hirano21} because some physical quantities related to the amplification or accumulation of magnetic field (or flux), such as the mass-to-flux ratio and kinetic energy, can be conserved.
The sink particle technique removes only the gas around the sink particles without removing magnetic flux.
Thus, the mass-to-flux ratio is not conserved and would decrease with time.
With a small mass-to-flux ratio, the magnetic flux is leaked out from the region around the sink particle, for example due to interchange instability \citep{Zhao11,Machida20}.
We also showed that rotation around protostars amplifies the magnetic field.
However, the rotational energy, which is proportional to the mass, is substantially removed with the sink particle technique.
In addition, the high-density gas is not coupled with the magnetic field after removing the gas.
Thus, the rotational amplification of the magnetic field should be underestimated with the sink particle technique.
For these reasons, we used the stiff-EOS technique.
It is expected that the difference in the results among recent studies (rare or frequent fragmentation) is attributed to the treatment of protostar (sink particle or stiff-EOS techniques) and the inclusion of turbulence.

Finally, we comment on observational constraints of the first stars.
Our result is consistent with the observational constraint with no observation of low-mass ($<\!0.8\,\msun$) surviving first stars in the Galaxy.
Though the amplified magnetic field prohibits small-mass disk fragmentation, it is unclear whether the magnetic field affects wide binary/multiple \citep[separation $>10^3$\,au;][]{sugimura20} and chemo-thermal instability at the Jeans-scale \citep{hirano18sv}, which form more massive fragments.
If a massive first star binary forms from them, it could leave the massive star binary, which can be a promising progenitor of BH-BH merger like the gravitational wave sources \citep{kinugawa14}.
We are interested in the contribution of the amplified magnetic field to the contraction of the binary orbit, but it is outside the scope of this study.

\section{Conclusion} \label{sec:con}

We introduce a new exponential amplification mechanism of the magnetic field during the accretion phase of the first star formation.
Even if the star-forming gas cloud has only the cosmological magnetic field strength, the orbital rotation around the protostar amplifies the tiny magnetic seed to kilo-gauss as the current protostar in less than ten years after the protostar formation.
The amplified magnetic field region expands during the accretion phase and reaches $\sim\!10^4$\,au (inside which $\sim\!500\,\msun$) at $t_{\rm ps} \sim 10^5$\,yr when the protostar becomes the zero-age main-sequence.
Since the strong magnetic field completely prevents disk fragmentation, only one protostar forms in each accretion disk.
We conclude that the first star formation is inevitably affected by magnetic fields even if the initial magnetic field strength is a cosmological value, about $10^{-18}$\,G.

\cite{hirano21} showed the magnetic field amplification in the atomic hydrogen (H) cooling gas clouds.
This letter shows that the same amplification also occurs in molecular hydrogen (H$_2$) cooling gas clouds with lower accretion rates and a limited number of fragments.
The following Paper II will discuss in detail how the effects on gas cloud evolution depend on the initial magnetic field strength.
We note that the magnetic field amplification shown in this study would not operate in contemporary star formation because the magnetic field significantly dissipates within the disk. For this reason, we have overlooked this mechanism until today.

In the future, we will perform a parameter survey of the MHD simulations for the parameter ranges of the primordial star-forming gas clouds obtained from the cosmological simulations \citep{hirano14,hirano15}.
Although disk fragmentation is wholly eliminated in this study, we will check whether gas clouds with different physical parameters, such as accretion rate and rotation degree, result in the same or not.
In addition, the current simulations end at $t_{\rm ps} = 1000$\,yr, and additional calculations are needed to determine the final stellar mass at $t_{\rm ps} \sim 10^5$\,yr when the first star reaches the zero main sequence stage.
In the future, we will fully update the first star formation theory to incorporate MHD effects and determine the formation rates of observational counterparts, such as low-mass surviving stars and massive BH binaries.

\begin{acknowledgments}
This work used the computational resources of the HPCI system provided by the supercomputer system SX-Aurora TSUBASA at Cyber Sciencecenter, Tohoku University and Cybermedia Center, Osaka University through the HPCI System Research Project (Project ID: hp210004 and hp220003), and Earth Simulator at JAMSTEC provided by 2021 and 2022 Koubo Kadai.
S.H. was supported by JSPS KAKENHI Grant Numbers JP21K13960 and JP21H01123 and Qdai-jump Research Program 02217.
M.N.M. was supported by JSPS KAKENHI Grant Numbers JP17K05387, JP17KK0096, JP21K03617, and JP21H00046 and University Research Support Grant 2019 from the National Astronomical Observatory of Japan (NAOJ).
\end{acknowledgments}




\bibliography{ms}{}

\begin{thebibliography}{}
\expandafter\ifx\csname natexlab\endcsname\relax\def\natexlab#1{#1}\fi
\providecommand{\url}[1]{\href{#1}{#1}}
\providecommand{\dodoi}[1]{doi:~\href{http://doi.org/#1}{\nolinkurl{#1}}}
\providecommand{\doeprint}[1]{\href{http://ascl.net/#1}{\nolinkurl{http://ascl.net/#1}}}
\providecommand{\doarXiv}[1]{\href{https://arxiv.org/abs/#1}{\nolinkurl{https://arxiv.org/abs/#1}}}

\bibitem[{{Federrath} {et~al.}(2011){Federrath}, {Sur}, {Schleicher},
  {Banerjee}, \& {Klessen}}]{federrath11}
{Federrath}, C., {Sur}, S., {Schleicher}, D.~R.~G., {Banerjee}, R., \&
  {Klessen}, R.~S. 2011, \apj, 731, 62, \dodoi{10.1088/0004-637X/731/1/62}

\bibitem[{{Greif}(2015)}]{greif15}
{Greif}, T.~H. 2015, Computational Astrophysics and Cosmology, 2, 3,
  \dodoi{10.1186/s40668-014-0006-2}

\bibitem[{{Higuchi} {et~al.}(2018){Higuchi}, {Machida}, \& {Susa}}]{higuchi18}
{Higuchi}, K., {Machida}, M.~N., \& {Susa}, H. 2018, \mnras, 475, 3331,
  \dodoi{10.1093/mnras/sty046}

\bibitem[{{Hirano} \& {Bromm}(2017)}]{hirano17}
{Hirano}, S., \& {Bromm}, V. 2017, \mnras, 470, 898,
  \dodoi{10.1093/mnras/stx1220}

\bibitem[{{Hirano} \& {Bromm}(2018)}]{hirano18}
---. 2018, \mnras, 476, 3964, \dodoi{10.1093/mnras/sty487}

\bibitem[{{Hirano} {et~al.}(2015){Hirano}, {Hosokawa}, {Yoshida}, {Omukai}, \&
  {Yorke}}]{hirano15}
{Hirano}, S., {Hosokawa}, T., {Yoshida}, N., {Omukai}, K., \& {Yorke}, H.~W.
  2015, \mnras, 448, 568, \dodoi{10.1093/mnras/stv044}

\bibitem[{{Hirano} {et~al.}(2014){Hirano}, {Hosokawa}, {Yoshida}, {Umeda},
  {Omukai}, {Chiaki}, \& {Yorke}}]{hirano14}
{Hirano}, S., {Hosokawa}, T., {Yoshida}, N., {et~al.} 2014, \apj, 781, 60,
  \dodoi{10.1088/0004-637X/781/2/60}

\bibitem[{{Hirano} {et~al.}(2021){Hirano}, {Machida}, \& {Basu}}]{hirano21}
{Hirano}, S., {Machida}, M.~N., \& {Basu}, S. 2021, \apj, 917, 34,
  \dodoi{10.3847/1538-4357/ac0913}

\bibitem[{{Hirano} {et~al.}(2018){Hirano}, {Yoshida}, {Sakurai}, \&
  {Fujii}}]{hirano18sv}
{Hirano}, S., {Yoshida}, N., {Sakurai}, Y., \& {Fujii}, M.~S. 2018, \apj, 855,
  17, \dodoi{10.3847/1538-4357/aaaaba}

\bibitem[{{Hosokawa} {et~al.}(2010){Hosokawa}, {Yorke}, \&
  {Omukai}}]{hosokawa10}
{Hosokawa}, T., {Yorke}, H.~W., \& {Omukai}, K. 2010, \apj, 721, 478,
  \dodoi{10.1088/0004-637X/721/1/478}

\bibitem[{{Ichiki} {et~al.}(2006){Ichiki}, {Takahashi}, {Ohno}, {Hanayama}, \&
  {Sugiyama}}]{ichiki06}
{Ichiki}, K., {Takahashi}, K., {Ohno}, H., {Hanayama}, H., \& {Sugiyama}, N.
  2006, Science, 311, 827, \dodoi{10.1126/science.1120690}

\bibitem[{{Inayoshi} {et~al.}(2020){Inayoshi}, {Visbal}, \&
  {Haiman}}]{inayoshi20}
{Inayoshi}, K., {Visbal}, E., \& {Haiman}, Z. 2020, \araa, 58, 27,
  \dodoi{10.1146/annurev-astro-120419-014455}

\bibitem[{{Johns-Krull}(2007)}]{johns-krull07}
{Johns-Krull}, C.~M. 2007, \apj, 664, 975, \dodoi{10.1086/519017}

\bibitem[{{Kinugawa} {et~al.}(2014){Kinugawa}, {Inayoshi}, {Hotokezaka},
  {Nakauchi}, \& {Nakamura}}]{kinugawa14}
{Kinugawa}, T., {Inayoshi}, K., {Hotokezaka}, K., {Nakauchi}, D., \&
  {Nakamura}, T. 2014, \mnras, 442, 2963, \dodoi{10.1093/mnras/stu1022}

\bibitem[{{Koh} {et~al.}(2021){Koh}, {Abel}, \& {Jedamzik}}]{koh21}
{Koh}, D., {Abel}, T., \& {Jedamzik}, K. 2021, \apjl, 909, L21,
  \dodoi{10.3847/2041-8213/abe8dd}

\bibitem[{{Lazarian} {et~al.}(2020){Lazarian}, {Eyink}, {Jafari}, {Kowal},
  {Li}, {Xu}, \& {Vishniac}}]{lazarian20}
{Lazarian}, A., {Eyink}, G.~L., {Jafari}, A., {et~al.} 2020, Physics of
  Plasmas, 27, 012305, \dodoi{10.1063/1.5110603}

\bibitem[{{Lazarian} \& {Vishniac}(1999)}]{lazarian99}
{Lazarian}, A., \& {Vishniac}, E.~T. 1999, \apj, 517, 700,
  \dodoi{10.1086/307233}

\bibitem[{{Machida}(2014)}]{machida14}
{Machida}, M.~N. 2014, \apjl, 796, L17, \dodoi{10.1088/2041-8205/796/1/L17}

\bibitem[{{Machida} \& {Basu}(2020)}]{Machida20}
{Machida}, M.~N., \& {Basu}, S. 2020, \mnras, 494, 827,
  \dodoi{10.1093/mnras/staa672}

\bibitem[{{Machida} \& {Doi}(2013)}]{machida13}
{Machida}, M.~N., \& {Doi}, K. 2013, \mnras, 435, 3283,
  \dodoi{10.1093/mnras/stt1524}

\bibitem[{{Machida} \& {Nakamura}(2015)}]{machida15}
{Machida}, M.~N., \& {Nakamura}, T. 2015, \mnras, 448, 1405,
  \dodoi{10.1093/mnras/stu2633}

\bibitem[{{Magg} {et~al.}(2018){Magg}, {Hartwig}, {Agarwal}, {Frebel},
  {Glover}, {Griffen}, \& {Klessen}}]{magg18}
{Magg}, M., {Hartwig}, T., {Agarwal}, B., {et~al.} 2018, \mnras, 473, 5308,
  \dodoi{10.1093/mnras/stx2729}

\bibitem[{{McKee} {et~al.}(2020){McKee}, {Stacy}, \& {Li}}]{mckee20}
{McKee}, C.~F., {Stacy}, A., \& {Li}, P.~S. 2020, \mnras, 496, 5528,
  \dodoi{10.1093/mnras/staa1903}

\bibitem[{{Omukai} {et~al.}(2008){Omukai}, {Schneider}, \& {Haiman}}]{omukai08}
{Omukai}, K., {Schneider}, R., \& {Haiman}, Z. 2008, \apj, 686, 801,
  \dodoi{10.1086/591636}

\bibitem[{{Prole} {et~al.}(2022){Prole}, {Clark}, {Klessen}, {Glover}, \&
  {Pakmor}}]{Prole22}
{Prole}, L., {Clark}, P., {Klessen}, R., {Glover}, S., \& {Pakmor}, R. 2022,
  arXiv e-prints, arXiv:2206.11919.
\newblock \doarXiv{2206.11919}

\bibitem[{{Sadanari} {et~al.}(2021){Sadanari}, {Omukai}, {Sugimura},
  {Matsumoto}, \& {Tomida}}]{sadanari21}
{Sadanari}, K.~E., {Omukai}, K., {Sugimura}, K., {Matsumoto}, T., \& {Tomida},
  K. 2021, \mnras, 505, 4197, \dodoi{10.1093/mnras/stab1330}

\bibitem[{{Sharda} {et~al.}(2020){Sharda}, {Federrath}, \&
  {Krumholz}}]{sharda20}
{Sharda}, P., {Federrath}, C., \& {Krumholz}, M.~R. 2020, \mnras, 497, 336,
  \dodoi{10.1093/mnras/staa1926}

\bibitem[{{Sharda} {et~al.}(2021){Sharda}, {Federrath}, {Krumholz}, \&
  {Schleicher}}]{sharda21}
{Sharda}, P., {Federrath}, C., {Krumholz}, M.~R., \& {Schleicher}, D. R.~G.
  2021, \mnras, 503, 2014, \dodoi{10.1093/mnras/stab531}

\bibitem[{{Stacy} {et~al.}(2022){Stacy}, {McKee}, {Lee}, {Klein}, \&
  {Li}}]{stacy22}
{Stacy}, A., {McKee}, C.~F., {Lee}, A.~T., {Klein}, R.~I., \& {Li}, P.~S. 2022,
  \mnras, 511, 5042, \dodoi{10.1093/mnras/stac372}

\bibitem[{{Sugimura} {et~al.}(2020){Sugimura}, {Matsumoto}, {Hosokawa},
  {Hirano}, \& {Omukai}}]{sugimura20}
{Sugimura}, K., {Matsumoto}, T., {Hosokawa}, T., {Hirano}, S., \& {Omukai}, K.
  2020, \apjl, 892, L14, \dodoi{10.3847/2041-8213/ab7d37}

\bibitem[{{Sur} {et~al.}(2012){Sur}, {Federrath}, {Schleicher}, {Banerjee}, \&
  {Klessen}}]{sur12}
{Sur}, S., {Federrath}, C., {Schleicher}, D.~R.~G., {Banerjee}, R., \&
  {Klessen}, R.~S. 2012, \mnras, 423, 3148,
  \dodoi{10.1111/j.1365-2966.2012.21100.x}

\bibitem[{{Sur} {et~al.}(2010){Sur}, {Schleicher}, {Banerjee}, {Federrath}, \&
  {Klessen}}]{sur10}
{Sur}, S., {Schleicher}, D.~R.~G., {Banerjee}, R., {Federrath}, C., \&
  {Klessen}, R.~S. 2010, \apjl, 721, L134, \dodoi{10.1088/2041-8205/721/2/L134}

\bibitem[{{Susa}(2019)}]{susa19}
{Susa}, H. 2019, \apj, 877, 99, \dodoi{10.3847/1538-4357/ab1b6f}

\bibitem[{{Turk} {et~al.}(2012){Turk}, {Oishi}, {Abel}, \& {Bryan}}]{turk12}
{Turk}, M.~J., {Oishi}, J.~S., {Abel}, T., \& {Bryan}, G.~L. 2012, \apj, 745,
  154, \dodoi{10.1088/0004-637X/745/2/154}

\bibitem[{{Yoon} {et~al.}(2012){Yoon}, {Dierks}, \& {Langer}}]{yoon12}
{Yoon}, S.~C., {Dierks}, A., \& {Langer}, N. 2012, \aap, 542, A113,
  \dodoi{10.1051/0004-6361/201117769}

\bibitem[{{Zhao} {et~al.}(2011){Zhao}, {Li}, {Nakamura}, {Krasnopolsky}, \&
  {Shang}}]{Zhao11}
{Zhao}, B., {Li}, Z.-Y., {Nakamura}, F., {Krasnopolsky}, R., \& {Shang}, H.
  2011, \apj, 742, 10, \dodoi{10.1088/0004-637X/742/1/10}

\end{thebibliography}
\bibliographystyle{aasjournal}

\end{document}